# Information Analysis Infrastructure for Diagnosis

Vladimir Hahanov[1], Wajeb Gharibi[2], Eugenia Litvinova[1], Svetlana Chumachenko[1]

[1] Computer Engineering Faculty, Kharkov National University of Radioelectronics, Kharkov, Ukraine, hahanov@kture.kharkov.ua

[2] Kingdom of Saudi Arabia, Jazan University, gharibiw2002@yahoo.com

**Abstract**—A high-speed multiprocessor architecture for brain-like analyzing information represented in analytic, graph- and table forms of associative relations to search, recognize and make a decision in n-dimensional vector discrete space is offered. Vector-logical process models of actual applications, where the quality of solution is estimated by the proposed integral non-arithmetical metric of the interaction between binary vectors, are described. The theoretical proof of the metric for a vector logical space and the quality criteria for estimating solutions is created.

## 1. INTRODUCTION

The goal is to remove arithmetic from computer and transform free resources to the brain-like infrastructure of associative logic simulating the brain functionality that makes possible making the right decision every moment. The brain and the computer have the same technological basis in the form of primitive logical operations: and, or, not, xor. With experience, the brain and the computer create more complex functional space-time logic converters using the above primitive operations. Specialization of computer, focused on using only logical operations, enables to approximate to the associative logic human thinking, and thus considerably (x100) improve the performance of solving nonarithmetic problems.

Removing arithmetic operations, leveraging the parallelism of the vector logic algebra, and multiprocessor architecture provide an efficient infrastructure, which combines mathematical and technological culture to solve applied problems.

Brain-likeness of multiprocessor digital system-on-a-chip is the concept of making architecture and models of computational processes to implement typical brain nonarithmetic associative logic functionalities on today's digital platform by using vector logical operations and criteria for search, pattern recognition and decision-making problems. Market appeal of logical associative multiprocessor (LAMP) is determined by thousands of old and new logical problems, which now are solved ineffectively by redundant universal computers with high-performance arithmetic processor. Here are some problems relevant to the IT-market: 1. Analysis and synthesis of syntactic and semantic language structures (abstracting, error correction, analysis of the text quality). 2. Video and audio pattern recognition by means of their representation by vector models of essential parameters in discrete space. 3. Use of Infrastructure IP for complex technical products to ensure their manufacturability and lifetime reliability. 4. Knowledge testing and expert appraisal of objects or parties to determine their validity. 5. Identification of the object or process to make a decision under uncertainty. 6. Exact information retrieval in the Internet, if information is given by a vector of parameters. 7. Target designation of fighter or aircraft autoland system functioning in microsecond time. 8. Air traffic control or optimization of municipal traffic control infrastructure to avoid conflicts. Practically all these problems are solved in real time; they are isomorphic by the logical structure of the process models, based on a set of interrelated associative tables. To solve them it is necessary quick and dedicated hardware platform (LAMP), focused on the concurrent execution of search, recognition and decision-making procedures, estimated by means of the integral nonarithmetic quality criterion.

Our goal in this article is to increase considerably (x100) the speed of search, recognition and decision-making procedures by means of multiprocessor and concurrent implementation of associative logic vector operations for the analyzing graph and tabular data structures in discrete Boolean space without the use of arithmetic operations.

The problems are: 1) Developing nonarithmetic metric for estimating the associative logic solutions. 2) Creation of data structures and process models for solving the applied problems. 3) Designing the architecture of logical associative multiprocessor. 4) Implementation of LAMP.

References: 1. Hardware platform for associative logical information analysis [1-2]. 2. Associative logical data structures for solving the information problems [3-4]. 3. Models and methods for discrete analyzing and synthesizing [5-6]. 4. Multiprocessors for solving information-logical problems [7-10]. 5. Brain-like and intelligent logical computing [11-12].

## 2. INTEGRAL METRIC FOR SOLUTION ESTIMATION

The infrastructure of brain-like multiprocessor includes models, methods and associative logical data structure, focused on hardware implementation of searching, recognition and decision-making [11-12] on the basis of vector nonarithmetic operations.

Evaluation of problem solution is determined by the vector-logical criterion of interaction quality between a query (a vector m) and a system of associative vectors (associators). The query processing results in generating a positive response in the form of one or more associators, as well as the numerical grade of membership characteristic (quality function) of an input vector m to the obtained solution: $\mu(m \in A)$. The input vector $m = (m_1, m_2, \ldots m_i, \ldots m_q)$, $m_i \in \{0,1,x\}$ and the matrix $A_i$ of associators $A_{ijr} (\in A_{ij} \in A_i \in A) = \{0,1,x\}$ have the same dimension that is equal to $q$. Below the membership grade of m-vector to A is designated by $\mu(m \in A)$.

There are 5 types of set-theoretic (logical) $\Delta$ - interaction of two vectors $m \cap A$ defined in Fig. 1. They form all primitive reactions of the generalized SRM systems (SRM – Search, Pattern Recognition and Decision Making) on the input request vector. In the technological field of knowledge – Design & Test – this sequence of actions is isomorphic to the route: fault finding, fault locating, decision-making for repairing. All three stages of technological route require the metric for estimating solutions to choose the optimal variant.

Integral set-theoretic metric for the estimating quality of the query execution is interaction function of multivalued vectors $m \cap A$, which is determined by average sum of three normalized parameters: code distance $d(m, A)$, membership function $\mu(m \in A)$ and reverse

membership function $\mu(A \in m)$:

$$Q = \frac{1}{3}[d(m,A) + \mu(m \in A) + \mu(A \in m)],$$

$$d(m,A) = \frac{1}{n}[n - \text{card}(m_i \bigcap_{i=1}^{n} A_i = \varnothing)];$$

$$\mu(m \in A) = 2^{\text{card}(m \cap A) - \text{card}(A)} \leftarrow \text{card}(m \cap A) =$$
$$= \text{card}(m_i \bigcap_{i=1}^{n} A_i = x) \,\&\, \text{card}(A) = \text{card}(\bigcup_{i=1}^{n} A_i = x); \qquad (1)$$

$$\mu(A \in m) = 2^{\text{card}(m \cap A) - \text{card}(m)} \leftarrow \text{card}(m \cap A) =$$
$$= \text{card}(m_i \bigcap_{i=1}^{n} A_i = x) \,\&\, \text{card}(m) = \text{card}(\bigcup_{i=1}^{n} m_i = x).$$

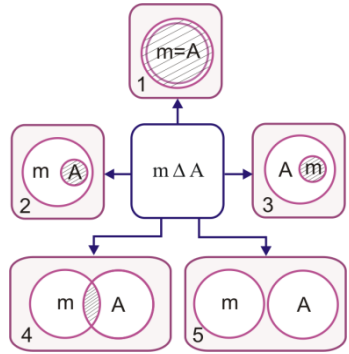

Fig. 1. The results of the intersection of two vectors

The normalization of parameters (variables) makes it possible to estimate the level of vector interaction within the interval [0,1]. The interacting vectors are equal if fixed maximum value of each variable equal to 1. The minimal estimation $Q = 0$ fixed if the vectors are not coincided by all n coordinates. If intersection power $m \cap A = m$ is equal to half of A-vector space, membership and quality functions are equal respectively:

$$\mu(m \in A) = \frac{1}{2};\ \mu(A \in m) = 1;\ d(m,A) = 1;$$
$$Q(m,A) = \frac{5}{2 \times 3} = \frac{5}{6}.$$

The same value will be set for Q parameter if the power of intersection $m \cap A = A$ is equal to half of the vector space m. If the power of intersection $\text{card}(m \cap A)$ is equal to half of the A-vector spaces power m, membership functions are the following:

$$\mu(m \in A) = \frac{1}{2};\ \mu(A \in m) = \frac{1}{2};\ d(m,A) = 1;$$
$$Q(m,A) = \frac{4}{2 \times 3} = \frac{4}{6} = \frac{2}{3}.$$

It should be noted, if the intersection between two vectors is equal to empty set, then the expression 2 power empty set is equal to zero: $2^{\text{card}(m \cap A) = \varnothing} = 2^{\varnothing} = 0$. It really means that the number of common points of two intersected spaces is zero.

The aim of a new vector logical solution criterion is improving considerably the performance of calculating the quality Q of interaction between the components m and A, when analysis associative data structures by using vector logical operations only. The arithmetic criterion (1) without the averaging membership functions and code distance can be transformed to the form:

$$Q = d[m, A_{i(j)}] + \mu[m \in A_{i(j)}] + \mu[A_{i(j)} \in m],$$
$$d(m, A_{i(j)}) = card[m \underset{i(j)=1}{\overset{n(m)}{\oplus}} A_{i(j)} = 1];$$
$$\mu(m \in A_{i(j)}) = card[A_{i(j)} = 1] - card[m \underset{i(j)=1}{\overset{n(m)}{\wedge}} A_{i(j)} = 1]; \quad (2)$$
$$\mu(A_{i(j)} \in m) = card[m = 1] - card[m \underset{i(j)=1}{\overset{n(m)}{\wedge}} A_{i(j)} = 1].$$

The first component of the criterion forms degree of mismatch between n-dimensional vectors so called the code distance, by performing xor operation. Second and third ones determine the degree of non-membership of conjunction result to a set of 1 for each of two interacting vectors. The notions of membership and non-membership are complementary, but calculating non-membership is more technological. Thus, the ideal criterion of quality is equal to zero, if two vectors are equal. The estimation of interaction quality between two binary vectors is decreasing with increasing criterion from 0 up to 1. To finally remove arithmetic operations, when counting a vector quality criterion, it is necessary to transform the expressions (2) to the form:

$$Q = d(m, A) \vee \mu(m \in A) \vee \mu(A \in m),$$
$$d(m, A) = m \oplus A;$$
$$\mu(m \in A) = A \wedge \overline{m \wedge A}; \quad (3)$$
$$\mu(A \in m) = m \wedge \overline{m \wedge A}.$$

Here the criteria are not numbers, but vectors, which determine the interaction between components m, A. The increasing 0-value quantity indicates improving vectors criterion quality, and 1's indicate loss of interaction quality. To compare the estimations it is necessary to determine the power of 1's in each vector without performing addition operation. This can be done by using the register [9-10] (Fig. 2), which makes it possible to perform left shifting and compacting all 1 coordinates of n-bit binary vector for one clock cycle.

After compacting procedure right unit bit number of compacted 1-set determines the index of interaction quality for vectors. For example, binary sets

$$m = (110011001100), A = (000011110101)$$

determining their interaction quality by formulas (3) is shown in the following form (zero coordinates are marked by dots):

| | |
|---|---|
| m | 1 1 . . 1 1 . . 1 1 . . |
| A | . . . . 1 1 1 1 . 1 . 1 |
| $m \wedge A$ | . . . . 1 1 . . . 1 . . |
| $\overline{m \wedge A}$ | 1 1 1 1 . . 1 1 1 . 1 1 |
| $d(m, A) = m \oplus A$ | 1 1 . . . . 1 1 1 . 1 1 |
| $\mu(A \in m) = m \wedge \overline{m \wedge A}$ | 1 1 . . . . . . 1 . . . |
| $\mu(m \in A) = A \wedge \overline{m \wedge A}$ | . . . . . . 1 1 . . . 1 |
| $Q = d(m, A) \vee \mu(m \in A) \vee \mu(A \in m)$ | 1 1 . . . . 1 1 1 . 1 1 |
| $Q(m, A) = (6/12)$ | 1 1 1 1 1 1 . . . . . . |

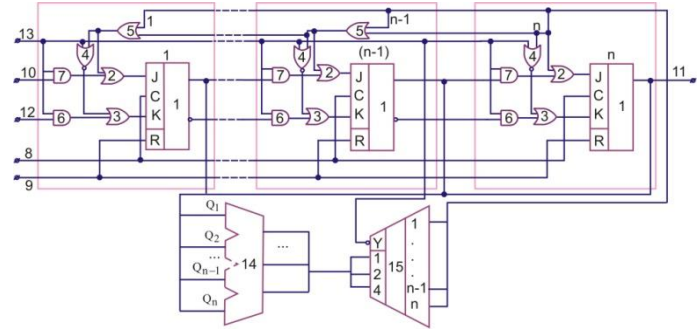

Fig. 2. Register for left shifting and compacting 1's

It is formed not only the estimation of vector interaction that is equal to $Q(m, A) = (6/12)$, but the most importantly, 1-unit coordinates of the row $Q = d(m, A) \vee \mu(m \in A) \vee \mu(A \in m)$ identify all essential variables for which there is local low-quality vector interaction. To compare two solutions obtained by logical analysis, compressed quality vectors $Q$ are used; and procedure including the following vector operations is performed:

$$Q(m,A) = \begin{cases} Q_1(m,A) \leftarrow \text{or}[Q_1(m,A) \wedge Q_2(m,A) \oplus Q_1(m,A)] = 0; \\ Q_2(m,A) \leftarrow \text{or}[Q_1(m,A) \wedge Q_2(m,A) \oplus Q_1(m,A)] = 1. \end{cases} \quad (4)$$

Vector-bit or-operator of devectorization determines a binary bit solution based on application a logical OR operation from n bits of an essential variables vector quality criterion. A circuit design for decision

$$Q = \begin{cases} Q_1 \leftarrow Y = 0 \\ Q_2 \leftarrow Y = 1 \end{cases}$$

and math process-model include 3 operations, shown in Fig. 3.

For binary vectors which have quality criteria, the procedure for choosing the best one based on the expression (4) is presented below:

| | |
|---|---|
| $Q_1(m,A) = (6,12)$ | 1 1 1 1 1 1 . . . . . . |
| $Q_2(m,A) = (8,12)$ | 1 1 1 1 1 1 1 1 . . . . |
| $Q_1(m,A) \wedge Q_2(m,A)$ | 1 1 1 1 1 1 . . . . . . |
| $Q_1(m,A) \oplus Q_1(m,A) \wedge Q_2(m,A)$ | . . . . . . . . . . . . |
| $Q(m,A) = Q_1(m,A)$ | 1 1 1 1 1 1 . . . . . . |

Vector logical criteria of interaction quality for associative sets is enable to obtain estimation of the search, pattern recognition and decision-making with high-speed parallel logic operations, which is especially important for critical real-time systems.

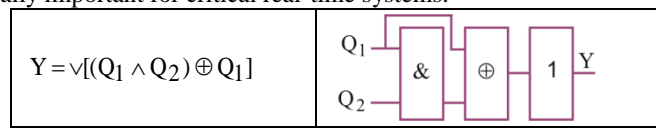

Fig. 3. Process-model of decision

The quality criterion Q uniquely determines three forms of interaction between any two objects in the n-dimensional vector logical space: the distance, and two membership functions. Taking into account that all three estimates included in the integral criterion form are joined by the function OR, simplification of vector interaction gives the result

$$Q = d(m, A) \vee \mu(m \in A) \vee \mu(A \in m) =$$
$$= m \oplus A \vee A \wedge \overline{m \wedge A} \vee m \wedge \overline{\overline{m} \wedge A} =$$
$$= m \oplus A \vee [A \wedge (\overline{m} \vee \overline{A})] \vee [m \wedge (\overline{\overline{m}} \vee \overline{A})] =$$
$$= m \oplus A \vee [A\overline{m} \vee A\overline{A} \vee m\overline{m} \vee m\overline{A}] =$$
$$= (A\overline{m} \vee m\overline{A}) \vee [A\overline{m} \vee A\overline{A} \vee m\overline{m} \vee m\overline{A}] =$$
$$= A\overline{m} \vee m\overline{A} \vee A\overline{m} \vee A\overline{A} \vee m\overline{m} \vee m\overline{A} =$$
$$= m \oplus A.$$

The quality criterion $Q = m \oplus A$ conforms to the metric for estimating distance or interaction in vector logical space, as well as it has a trivial computational form for estimating many solutions related to the analysis and synthesis of information. In fact, a logical vector space should not use the metric distance and quality criteria, including scalar arithmetic operations. Vector logic criterion (VLC) determines not only the distance between disjoint objects of a vector-logical space, but also their mutual membership: $d(a,b) \vee \mu(a,b) \vee \mu(b,a)$, if they overlap.

Metric $\beta$ of vector logic space is defined by a single equality that forms zero xor-sum of the distances between nonzero and finite quantity of points, closed in a cycle:

$$\beta = \bigoplus_{i=1}^{n} d_i = 0.$$

The metric $\beta$ of vector logical space is focused not on elements of the set, but the relationship, thereby reducing the axioms from three to one formula and extend it to arbitrary complex structures of n-dimensional space.

## 3. PROCESS MODEL FOR SEARCHING, RECOGNITION AND DECISION MAKING

The analytical form of a generalized process model for choosing the best interaction between the input query m and the system of logic associative relations is presented in the form:

$$P(m, A) = \min Q_i (m \underset{i=1}{\overset{n}{\Delta}} A_i) = \vee[(Q_i \underset{j=1,n}{\overset{j \neq i}{\wedge}} Q_j) \oplus Q_i] = 0;$$
$$Q(m, A) = (Q_1, Q_2, \ldots Q_i, \ldots Q_n);$$
$$A = (A_1, A_2, \ldots A_i, \ldots A_n);$$
$$\Delta = \{and, or, xor, not, slc, nop\};$$
$$A_i = (A_{i1}, A_{i2}, \ldots A_{ij}, \ldots A_{is});$$
$$A_{ij} = (A_{ij1}, A_{ij2}, \ldots A_{ijr}, \ldots A_{msq});$$
$$m = (m_1, m_2, \ldots m_r, \ldots m_q).$$
$$Q_i = d(m, A_i) \vee \mu(m \in A_i) \vee \mu(A_i \in m),$$
$$d(m, A_i) = m \oplus A_i;$$
$$\mu(m \in A_i) = A_i \wedge \overline{m \wedge A_i};$$
$$\mu(A_i \in m) = m \wedge \overline{\overline{m} \wedge A_i}.$$

Comment: 1) The functionality $P(m, A)$ specifies the analytical model for computational process in the form of statement, minimizing the integral quality criterion. 2) Data structures are presented as nodes-tables of the graph $A = (A_1, A_2, \ldots A_i, \ldots A_m)$, which logically interact each other. 3) A graph node is described by the ordered set of the vector-rows of an associative table $A_i = (A_{i1}, A_{i2}, \ldots A_{ij}, \ldots A_{is})$ for explicit solutions, where the row $A_{ij} = (A_{ij1}, A_{ij2}, \ldots A_{ijr}, \ldots A_{msq})$ is true proposition. Since the functional presented in tabular form has no time-constant input and output variables, this structure differs from sequential von Neumann's machine, defined by finite automata Miles and Moore. Equivalence of all variables in the vector $A_{ij} = (A_{ij1}, A_{ij2}, \ldots A_{ijr}, \ldots A_{msq})$ creates conditions for their existence that means the invariance of the problem solving for direct and inverse implication in the space $A_i \in A$. The associative vector $A_{ij}$ is an explicit solution, where each variable is defined in the final, multi-valued and discrete alphabet $A_{ijr} \in \{\alpha_1, \alpha_2, \ldots \alpha_i, \ldots \alpha_k\} = \beta$. The interaction $P(m, A)$ between the input vector-query $m = (m_1, m_2, \ldots m_r, \ldots m_q)$ and the graph $A = (A_1, A_2, \ldots A_i, \ldots A_m)$ generates a set of solutions and makes it possible to choose the best ones by minimum quality criterion:

$$P(m, A) = \min Q_i [m \wedge (A_1 \vee A_2 \vee \ldots \vee A_i \vee \ldots \vee A_m)].$$

The concrete interaction between the graph nodes generates the functionality $A = (A_1, A_2, \ldots A_i, \ldots A_m)$ that can be realized by the following structures: 1) A single associative table that includes all solutions of a logic problem explicitly. The advantage is maximum speed of parallel associative searching for a solution by the table. The disadvantage is the highest hardware complexity of memory allocation for large-scale table. 2) Tree (graph) structure of binary relations between the functional primitives, each of them generates the truth table for small numbers of variables. The advantage is the smallest hardware complexity of problem solving. The disadvantage is minimum speed of sequential associative searching for a solution by tree. 3) The compromise graph structure of logically understandable to the user relations between primitives, each of them generates the truth table for logical strongly connected variables. The advantage is high speed of concurrent associative searching for solutions by minimal number of the graph tables, as well as relatively low hardware complexity of problem solving. The disadvantage is decrease in speed because of sequential logic processing of the graph structure for explicit solutions found in the tables.

Partitioning a single table (associative memory) on $k$ parts allows reducing hardware cost, expressed in components (LUTs – Look Up Table) of programmable logic array [8,9].

The proposed process model for analyzing the graph of associative tables, as well as the introduced solution quality criteria are the basis of the developing a dedicated multiprocessor architecture focused on the concurrent vector logic operations.

## 4. ARCHITECTURE OF LOGIC ASSOCIATIVE MULTIPROCESSOR

To analyze large information volumes of logical data, there are several technologies focused to the practical application: 1. Use a workstation for serial programming, where the cost and time of problem solving are very high. 2. Development of a dedicated concurrent processor based on the PLD. The high concurrency of information processing compensates for the relatively low clock rate in comparison with CPU. Such reprogrammable circuit design is the best solution regarding performance. Disadvantage is lack of flexibility the software methods for solving logic problems and high cost of implementing the system-on-a-chip PLD under large production volumes. 3. The best solution is to leverage advantages CPU, PLD and ASIC concurrently [8,9]. This is due to the flexibility of programming, the possibility of correcting the source code, the minimum command set, and simple circuit designs for hardware multiprocessor im-plementation, the parallelization of logic procedures by the structure of bit processors. The implementation of a multiprocessor in ASIC allows to obtain the maximum clock rate, the minimum chip cost for large product volumes, and low power consumption. Combining the advantages of the technologies above determines the basic configuration of the LAMP, which has spherical multiprocessor structure (Fig. 4), consisting of 16 vector sequencers. Each sequencer together with the boundary elements is connected with eight contiguous ones. The processor PRUS [9], developed by Dr. Stanley Hyduke (CEO Aldec, USA), is the LAMP prototype.

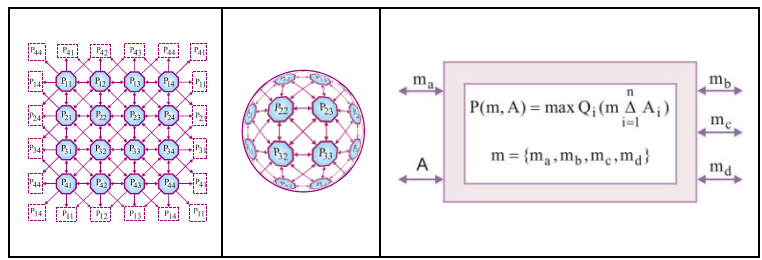

Fig. 4. LAMP macroarchitecture and interface

Entering information in the processor is realized like the classical design flow, except the stage "place and route" that is replaced by the operation of distributing software modules and data among all logical bit processors running concurrently. The compiler provides the placement of data among processors, sets the time of searching for solutions at the output each of them, and also plans transfer the results to another processor. LAMP is an effective processor network, which processes the data and provides the exchange of information between network components when searching for solution.

The simple circuit engineering of each processor can effectively process very large arrays with millions bits of information, expending time in hundreds the times less compared with general-purpose processor. Basic cell (vector processor for LAMP) can be synthesized by using 200 gates, which makes it possible to implement network containing 4096 computers in ASIC, using advanced silicone technology. Taking into account that memory costs for data storage are very small, LAMP may be applied for the designing of control systems in the areas of human activity, such as: industry, medicine, information protection, geology, weather forecasting, artificial intelligence, space science. LAMP is of particular interest for digital data processing, pattern recognition and cryptanalysis. If LAMP functioning is considered, its main purpose is obtaining quasi-optimal solution of the integrated problem of search and / or pattern recognition by using infrastructure components focused to the performing vector logical operations: $P(m, A) = \min Q_i (m \Delta A_i)$, $m = \{m_a, m_b, m_c, m_d\}$. System interface, corresponding to this functional, is presented in Fig. 2. All components $\{A, m_a, m_b, m_c, m_d\}$ can be input and output. Bidirectional interface specification is related to the invariance of relation for all variables, vectors, A-matrix, components and infrastructure inputs and / or outputs. Therefore, the structural model of LAMP can be used to solve any problems of direct and inverse implication in discrete logical space, and it emphasizes its difference from the automaton model concept of computer with explicit inputs and outputs. The components or registers $m = (m_a, m_b, m_c, m_d)$ are used for solution in the form of buffer, input and output vectors, as well as for identification of quality estimation for request performance. One of the variants multiprocessor architecture LAMP is a structure shown in Fig. 5. The main its component is a multiprocessor matrix

$P = [P_{ij}], \text{card}(4 \times 4)$, containing 16 vector-processors, each of them is designed for performing 5 logic vector operations with data memory contents, described by a table of dimension $A = \text{card}(m \times n)$.

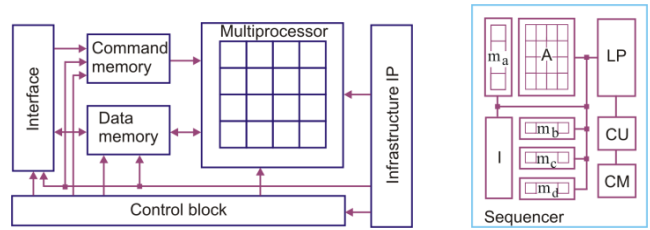

Fig. 5. LAMP architecture and sequencer structure

Interface is used for data exchange and data loading to the appropriate memory commands. The control unit initializes the executing commands of logical data processing and synchronizes the functioning all components of a multiprocessor.

Infrastructure IP [1] is designed for servicing all modules, diagnosing faults and repairing functionality of components and device in whole. Elementary logic associative processor or sequencer (see Fig. 5) is a part of the multiprocessor and contains: logical processor (LP), associative (memory) A-matrix for concurrent executing basic operations, block of vectors m, designed for concurrent processing rows and columns of A-matrix, as well as data exchange when computing, direct access memory (CM) for the storing commands of data processing software, automaton (CU) for logic operations execution control, interface (I) for the connecting sequencer and other elements of a multiprocessor. Logical Processor (LP) (Fig. 6) provides the implementation of five operations (and, or, not, xor, s1s - shift left bit crowding), which are the basis for the creating algorithms and procedures of information retrieval and evaluation of solutions.

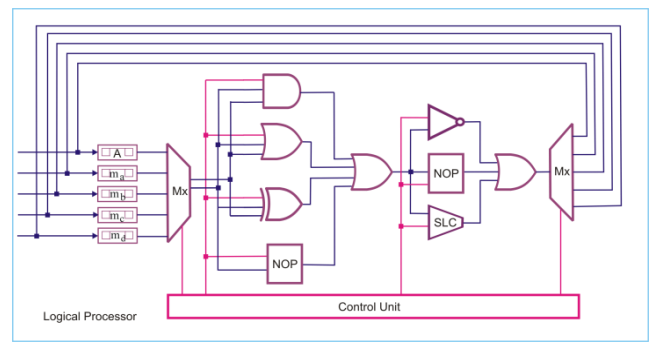

Fig. 6. Structure of logic calculations

LP module has a multiplexer at the input to select one of five operands, which is passed to the selected logic vector operator. By using a multiplexer (element or), a result is entered in one of four operands, which are selected by appropriate address.

Implementation features of the logical processor is use of three binary (and, or, xor) and two unary (not, slc) operations. The last ones can be added to the cycle of processing register data by selecting one of three operations (not, slc, nop – no-operation). To improve the efficiency of logical unit, two elements with empty operation are included. If it is necessary to perform a unary operation only, the selecting nop at the level of binary commands should be done, that almost means the transfer data through a follower to second level of unary operations. All LP op-erations are register or register-matrix. The last ones are designed for the analyzing vector-rows of a table using input m-vector as a request for exact information retrieval. The following combination of operators and operands are acceptable in a unit for logic calculating:

$$C = \begin{cases} \{m_a, m_b, m_c, m_d\}\Delta A_i; \\ \{m_a, m_b, m_c, m_d\}\Delta\{m_a, m_b, m_c, m_d\}; \\ \{not, nop, slc\}\{m_a, m_b, m_c, m_d, A_i\}. \end{cases}$$
$$\Delta = \{and, or, xor\}.$$

Realization of all vector operations for logic calculating by using a single sequencer in Verilog environment and followed implementation in PLD chip gives the results:

Logic Block Utilization: Number of 4 input LUTs: 400 out of 9,312 4%
Logic Distribution: Number of occupied Slices: 200 out of 4,656 4%
Number of Slices only related logic: 200 out of 200 100 %
Total Number of 4 input LUTs: 400 out of 9,312 4%
Number of bonded IOBs: 88 out of 320 29%
Total equivalent gate count for design: 2400

Clock rate of register operations for Xilinx's Virtex 4 is 100 MHz that by order of magnitude higher than similar procedures for a computer with clock rate 1GHz.

## 5. INFRASTRUCTURE FOR VECTOR LOGIC ANALYZING

Infrastructure is a set of models, methods and data definition languages, data analysis and synthesis tools for solving the functional problems. Model (system model) is a set of interrelated components defined in space and time, which describe the process or phenomenon with specified adequacy, and used for achieving the aim under constraints and metric for evaluating of the solution quality. Here, the constraints are the hardware costs, the time-to-market, which are have to be minimized. Metric for the evaluating solution by using the model is defined by a binary logic vector in the discrete Boolean space. The conceptual computer model is presented by an aggregate of control and operational automata. The system functionality model LAMP uses GALS (Global Asynhronus Local Synchronus) [8] technology for the creating hierarchical digital systems with the local synchronization of individual modules and simultaneous global asynchronous of the entire device.

To detail the structure of the vector processor and sequencer the analytical and structured process models are presented below. They are reduced to the analysis of the A-matrix by columns or rows. The first one is shown in Fig. 7 and it is designed for determining a set of feasible solutions relatively the input query $m_b$.

$$m_{ai}^m = \overline{\vee} \, [(m_b \underset{i=1}{\overset{n}{\wedge}} A_i) \oplus m_b];$$
$$A_i = (m_b \underset{i=1}{\overset{n}{\wedge}} A_i).$$

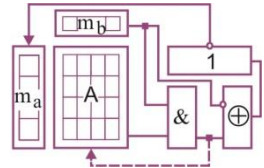

Fig. 7. Searching all feasible solutions

The second structure (Fig. 8) searches for optimal solution on a set of ones, found in the first process model, by analyzing rows. In addition, the second model has a separate application, focused on the finding single-valued and multi-valued solutions, for example, when searching for faults in digital systems-on-chips.

All operations are presented by two process models are vector. Process model for analyzing rows (see Fig. 7) generates the vector $m_a$ for identification of feasible $m_{ai}=1$ or contradictory $m_{ai}=0$ solutions relatively the input condition $m_b$ for n cycles of processing all m-bit vectors of the table $A = \text{card}(m \times n)$. The quality (validity) of the decision is determined for each interaction between the input vector $m_b$ and the row $A_i \in A$ by the disjunction (devectorization) block. The matrix $A$ can be modified by its intersecting with an input vector on the basis of the operation $A_i = (m_b \wedge \bigwedge_{i=1}^{n} A_i)$, if it is necessary to remove from the A-table all insignificant for the solution coordinates and vectors, marked by unit values of the vector $m_a$.

$$m_b^s = (\bigwedge_{\forall m_{ai}=1} A_i) \wedge \overline{(\bigvee_{\forall m_{ai}=0} A_i)}$$

$$m_b^m = (\bigvee_{\forall m_{ai}=1} A_i) \wedge \overline{(\bigvee_{\forall m_{ai}=0} A_i)}$$

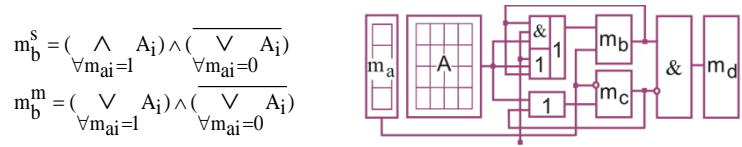

Fig. 8. Structure for searching the optimal solution

An interesting solution for the problems of diagnosis by analyzing table rows shown in Fig. 8, should be interpreted as follows. After performing the diagnostic experiment the binary output response vector $m_a$ is made, which masks the A-table of faults to detect single or multiple faults. Vectors $m_b$ and $m_c$ are used to accumulate the results of conjunction and disjunction operations. Then the logical subtraction the contents of the second vector $m_c$ from the first register $m_b$ and subsequent saving the result in register $m_d$ is performed. To implement the second equation, which generates a multiple solution, element AND is replaced by the function OR. The circuit has also a variable for the choosing the solution search mode: single or multiple. The process model uses as input condition a vector $m_a$, which controls the choice of vector operation AND, OR for processing unit $A_i(m_{ai}=1) \in A$ or zero $A_i(m_{ai}=0) \in A$ A-table rows. The result of n cycles is accumulation of unit and zero solutions relatively coordinate values of the vector $m_a$ in the registers $A_1, A_0$, respectively. A priori, the vectors of 1's and 0's are entered in these registers: $A_1 = 1, A_0 = 0$. After processing all n rows of A-table for n cycles the vector conjunction for the contents of register $A_1$ and the inversion of the register $A_0$ are performed, which generates the result in the form of the vector $m_b$, where unit coordinates determine a solution. When analyzing a fault table of digital device the columns identified with the numbers of faults or faulty blocks to be repaired correspond to unit coordinates of the vector $m_b$. Within the bounds of the Infrastructure IP the optimization repairing problem can be solved by using an universal structure of vector logic analysis. It is necessary to cover all faults found in the cells by minimum number of spare rows and/or columns, such as memory. The technological and mathematical culture of vector logic in this case provides a simple and interesting circuit solution for obtaining a quasioptimal coverage, Fig. 9. The advantages are: 1) The computational complexity of the procedure is $Z = n$ of vector operations, equal to the number of table rows. 2) The minimum hardware costs, which are a table and two vectors $m_b, m_a$ for storing intermediate coverages and the accumulating result in the form of unit coordinates, corresponding to table rows, which contain a quasioptimal coverage. 3) There is no need for the classical splitting the coverage problem for searching a coverage core and a complement. 4) There is no need for complicated procedures for manipulating rows and columns. The disadvantage is obtaining not always optimal coverage that is costs for the efficiency of vector procedure, shown in Fig. 9.

$$\begin{cases} m_b = (m_b \vee A_i); \\ m_{ai} = \overset{n}{\underset{i=1}{\vee}} [(m_b \vee A_i) \wedge \overline{m}_b]. \end{cases}$$

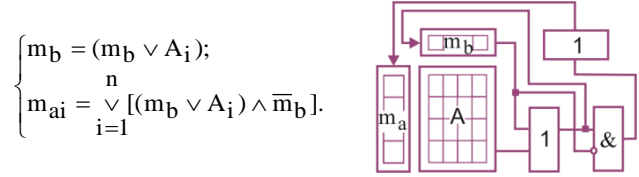

Fig. 9. Process model for searching quasioptimal coverage

There is the devectorization operation, which at the last stage transforms the vector result in a bit $m_{ai}$ of the vector $m_a$ by the function OR $m_{ai} = \vee[(m_b \vee A_i) \wedge \overline{m}_b]$. In general, in the algebra of vector operations the devectorization operation is written in the notation: <binary operation> <vector>: $\vee A_i, \wedge m, \overline{\wedge}(m \vee A_i)$. The inverse vectorization operation is the concatenation of Boolean variables: $m_a(a,b,c,d,e,f,g,h)$. In the process for coverage searching a priori the vectors $m_b = 0$, $m_a = 0$ are nulled. The quasioptimal coverage is accumulated in the vector $m_a$ for n cycles by serial shifting. Bits, entered in the register $m_a$, are formed by the circuit OR, which realizes devectorization by analyzing the input result $[(m_b \vee A_i) \wedge \overline{m}_b]$ on the presence of 1's. The next example is characterized by functionally completeness of the diagnosis cycle, when this information is used to repair faulty memory cells after obtaining the quasioptimal coverage [9]. The dimension of a memory module 13x15 cells does not influence on the computational complexity of obtaining a coverage for ten faulty cells by using spare rows (2) and columns (5) (Fig. 10).

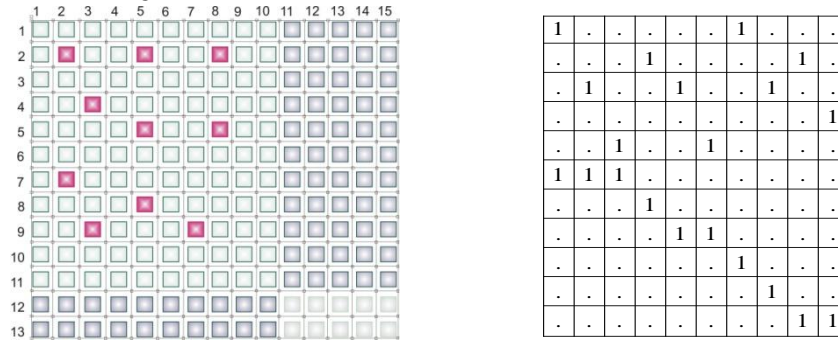

Fig. 10. Memory module with spare and coverage table

To solve the optimization problem a coverage table for faulty cells is generated (see Fig. 10); it contains rows (spares) for the covering faults: $(C_2, C_3, C_5, C_7, C_8, C_2, R_2, R_4, R_5, R_7, R_8, R_9)$. The columns are faults of cells $(F_{2,2}, F_{2,5}, F_{2,8}, F_{4,3}, F_{5,5}, F_{5,8}, F_{7,2}, F_{8,5}, F_{9,3}, F_{9,7})$ to be repaired. Here the columns match the coordinates of faulty cells, and rows identify the spare components (rows and columns), which can repair the faulty coordinates. The process model (Fig. 11) makes it possible to obtain the optimal solution in the form $m_a = \boxed{1\ 1\ 1\ 1\ 1\ 0\ 0\ 0\ 0\ 0}$, which corresponds to the coverage: $R = \{C_2, C_3, C_5, C_7, C_8\}$. It is one of three possible minimum solutions:

$$R = C_2, C_3, C_5, C_7, C_8 \vee C_2, C_3, C_5, C_8, R_9 \vee C_2, C_5, C_8, R_4, R_9$$

for a fault detection table. The technological model for embedded diagnosing and repairing memory is shown in Fig. 11.

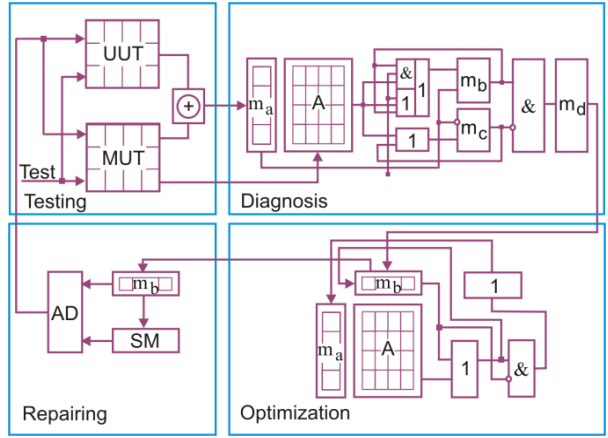

Fig. 11. Model for embedded testing and repairing memory

It includes four components: 1) Testing a module (UUT – Unit Under Test) by using the reference model (MUT – Model Under Test) to generate the output response vector $m_a$, dimension of which corresponds to the number of test patterns. 2) Fault diagnosis based on analysis of the fault detection table A. 3) Optimization of fault coverage by spares (rows and columns) based on the analysis of the table A. 4) Repairing memory by the readdressing (AD – Address Decoder) faulty rows and columns of the vector $m_a$ by spare components SM – Spare Memory [9].

The process model for embedded servicing operates in real time and allows maintaining a digital system-on-a chip without human intervention that is an interesting solution for the critical technologies related to the remote maintenance of a product. The proposed process model for the analysis associative tables, as well as the imposed quality criteria for logical solutions allows solving the problems for quasi-optimal covering, diagnosis software faults and/or hardware modules. The model of vector calculations provided the basis for the developing dedicated multiprocessor architecture focused to searching, pattern recognition and decision making by using associative tables.

The quality, Fig. 12, of design solution based on specialization Sp and standardization St requires use three discrepant parameters: yield Y, time-to-market T, hardware complexity H:

The parameter Y depends on design testability Q, the probability P of existence faulty areas in the chip and the number n of undetected faults.

$$E = F(L,T,H) = \min[\frac{1}{3}(L + T + H)],$$
$$Y = (1-P)^n;$$
$$L = 1 - Y^{(1-k)} = 1 - (1-P)^{n(1-k)};$$
$$T = \frac{(1-k) \times H^s}{H^s + H^a}; \quad H = \frac{H^a}{H^s + H^a}.$$

Fig. 13. Design solution quality for process model

The parameter L is a complement of the parameter Y (yield). It depends on the design testability k, the existence probability P of the faulty components and the quantity of undetected errors n. The verification time is determined by the design testability k, multiplied by the structur-al complexity of hardware-software functionality and divided by the total complexity of a design in the code lines or gates. Software-hardware redundancy depends on the complexity of the assertion code or boundary scan, divided by the total complexity of a design. At that assertion or boundary scan redundancy has to provide the specified diagnosis depth of the functionality errors during time-to-market, defined by customer.

## 6. CONCLUSION

Existing software analogs do not provide a purely vector-logical paths for searching, pattern recognition and decision-making in a discrete information spaces [3,8]. Almost all of them use a universal command system of modern expensive CPU with math coprocessor. On the other hand the hardware dedicated tools for logical analysis, which can be considered as prototypes [1,3], typically focused on bitwise or nonvector information processing.

To eliminate the disadvantages of software analogs and hardware prototypes it is proposed the new approach for vector logic processing the associative data with complete exclusion of arithmetic operations, which influence on the performance and hardware complexity. It was successfully implemented on the basis of modern microelectronic devices in the form of multiprocessor digital system-on-a-chip.

Actual implementation of the approach is based on the proposal of infrastructure, which includes the following components: 1. Process models for the analyzing associative tables based on the use of vector logical operations for searching, pattern recognition, decision making in the vector discrete Boolean space. Models are focused on high-performance concurrent vector logical analysis of information and calculating of solution quality criteria on the basis of proposed beta-metric of cyber space. 2. A multiprocessor architecture for concurrent solving associative logic problems by using a minimal set of vector logical operations and total exclusion of arithmetic instructions. It provides high performance, minimal cost and low power consumption of LAMP, implemented in a chip of programmable logic. 3. Novel vector logical process model for embedded diagnosing digital systems-on-chips and searching for quasioptimal coverage based on the logic associative multiprocessor, parallel operations for computing processes and calculating quality criteria.